# Using dopants as agents to probe key electronic properties of organic semiconductors

A. Fediai\*, F. Symalla, T. Neumann

Griesbachstr. 5, 76185 Karlsruhe, Germany

E-mail: artem.fediai@nanomatch.com

W. Wenzel

Institute of Nanotechnology, Karlsruhe Institute of Technology,

Hermann-von-Helmholtz-Platz 1, 76344, Eggenstein-Leopoldshafen, Germany

# Abstract

In organic electronics, conductivity doping is used primarily to eliminate charge injection barriers in OLEDs and other electronic devices. Therefore, research on conductivity doping was primarily focused on understanding and enhancing the properties of these doped layers. In contrast, this work shifts the focus from optimizing doped layers to leveraging the doping process as a tool for investigating fundamental material properties. Specifically, we use the dopant as an "agent" to enable the measurement of three critical parameters- ionization potential (IP), electron affinity (EA), and Coulomb interaction energy ($V_C$) - that govern dopant ionization and play central roles in organic electronic devices from OLEDs to OPVs in general. While these parameters can be measured experimentally, conventional approaches often involve intricate or indirect methods, such as spectral deconvolution, which may introduce ambiguities or fail to represent bulk properties. Here we show how consolidating the experimental data and simulations on the dopant ionization fraction and doped-induced conductivity can be used to estimate the mean IP of the embedded organic molecule, and $V_C$ of the embedded charge-transfer complex, each of the central importance in OLEDs or OPV design, respectively. Our results illustrate how measuring and simulating doped materials can provide access to the fundamental design parameters of organic electronic devices.

# 1. Introduction

The conductivity doping of organic semiconductors is crucial from multiple perspectives. In OLED technology, for instance, many commercially viable and efficient OLED architectures contain doped layers that reduce injection barriers and enhance the conductivity of electron or hole transport layers.[1–5] More broadly, doping is vital for any electronic device based on organic electronic materials. This is especially important for small-molecule organic semiconductors (OS), which typically exhibit orders of magnitude lower intrinsic mobility than their inorganic counterparts. The fundamental mechanism of doping in organic semiconductors differ from their inorganic counterparts, therefore experimental and theoretical efforts were mainly focused on understanding the doping mechanism, doping efficiency and structure-property relationships.[5–15] In contrast, this work considers doping from a different perspective: the doping efficiency is sensitive and clearly related to the ionization potential (IP) and electron affinity (EA) values of the host and dopant molecules embedded into doped layer. Embedded EA and IP values are crucial for every layer of organic electronic devices, such as the OLED stack, because they determine charge accumulation and depletion, injection barriers, charge carrier balance, and more. We note that IP and EA of molecules embedded within the organic material is very difficult and challenging experimentally due to several persistent issues: the role of energetic disorder (only onset values are measured), broadening due to experimental set up, and a penetration depth of 1–7 nm (depending on method), which means the measured values may partially reflect surface properties.[16] Spectroscopically measured IP and EA can also be strongly skewed if the material creates a surface potential due to spontaneous orientation polarization (SOP); as the vacuum level within the material differs strongly from without. In contrast, the charge transfer process between donor and acceptor in the bulk phase is less sensitive to these effects as the charge transfer is: a) unidirectional and b) occurs over short distances where the vacuum level shift is below 0.1 eV for most systems. Thus, doping can serve as an indirect method to determine the IP or EA in these materials.

In this work we combine experimental data with multi-scale simulations for *p*-doped small molecule organic materials to evaluate the possibility of extracting IP and $V_C$ from experimental data on doped layers. Experimental data of two kinds are collected from literature: (a) measured conductivity of materials doped by dopants of different strength and (b) measured dopant ionization fraction using UV-vis-NIR spectroscopy. In the simulation approach, we mimic physical vapor deposition to generate morphologies of doped thin films with atomistic resolution,[17] compute IP, EA and $V_C$ for various (pairs of) molecules taking into account the unique electrostatic environment of each molecule,[10,18] and use kMC to conduct "virtual experiments", i.e. to simulate charge transport and dynamics of the dopant ionization process to extract conductivity and dopant ionization fraction.[9,10]

The first part of the manuscript (**Section 2.1**) studies the dopant ionization fraction; the second part (**Section 2.2**) is devoted to the dopant EA - dependent conductivity. We demonstrate how using the synergy of experimental results reported for a range of *p*-doped materials and simulations of their "virtual" counterparts allows us to eliminate limitations of purely theoretical or experimental studies. Specifically, we show that the mean IP of the embedded host molecule may be indirectly measured with *chemical accuracy* by doping it with a range of variable-EA dopants. Additionally, we show that the embedded $V_C$ magnitude may be derived as the declination point on the plot where the conductivity of the doped material is plotted versus the IP(host)-EA(dopant), known as the energy offset.

# 2. Results and Discussion

## 2.1. Dopant Ionization.

### 2.1.1. Correlation with the offset IP-EA

We first analyze the experimental data available from the literature for UV-vis-NIR measurements for *p*-doped small molecule amorphous semiconductors. **Table 1** summarizes the fraction of the ionized p-dopants derived from experimental measurements in comparison with values computed using the multi-scale workflow. Note that in the experimental sources in literature additional materials were studied that are not included in this work due to their molecular size or their tendency to form (poly-)crystalline structures, both of which are difficult to tackle from a computational perspective.
**Figure 1** shows the whole multiscale workflow used in this work. Molecular geometries of all host and dopant molecules were optimized and intramolecular force-fields were computed using the DihedralParametrizer module.[19] Based on molecular structures and force-fields, 3D morphologies of thin films with atomistic resolution were generated by mimicking physical vapor deposition, as implemented in Deposit.[17] For each host-dopant combination, a morphology was generated at the respective molar fraction as specified in **Table 1**. Consequently, the QuantumPatch method was applied in three different modes to compute the electronic structure of molecules in the thin film phase:[10,18] 1) in the uncharged state of the molecules to determine the energetic disorder of every doped material 2) in the charged state of a eight molecules in the thin film to compute IP and EA, and 3) for 50 host-dopant pairs coulomb interaction energy $V_C$ was computed by computing charge transfer states as reported previously.[10] Based on these quantities, a kinetic Monte Carlo method was used to simulate the dynamics of host-dopant activation, i.e. the charge transfer between dopant and host, in order to compute the dopant ionization fraction.

Details of the specific settings used in simulations can be found in the Methods section.

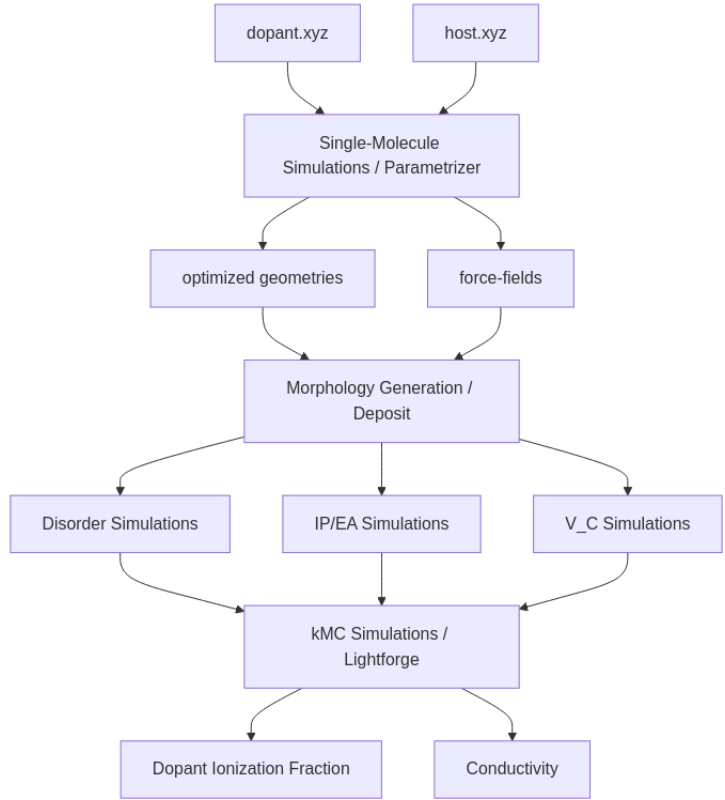

**Figure 1.** Multiscale Workflow used to simulate the dopant ionization fraction $\eta_{sim}$ and conductivity of *p*-doped materials employed in this work.

**Table 1.** Doped materials, dopant molar fraction in mol%, Experimental and simulated dopant ionization fractions ($\eta_{exp}$, $\eta_{sim}$). Relative dopant ionization fraction $\eta_{sim, rel}$ is only reported for materials from reference [a].

| | Material (<host>@<dopant>) | mol % | $\eta_{exp}$ | $\eta_{sim}$ | $\eta_{sim, rel}$ |
|---|---|---|---|---|---|
| 1 | BFDPB@CN6-CP[a] | 26 | 1.00 | 0.64/1.00 | 1.00 |
| 2 | NPB@CN6-CP[a] | 26 | 0.90 | 0.62/0.96 | 0.96 |
| 3 | BPAPF@CN6-CP[a] | 26 | 0.88 | 0.59/0.92 | 0.92 |
| 4 | TCTA@CN6-CP[a] | 26 | 0.42 | 0.54/0.84 | 0.84 |
| 5 | CBP@CN6-CP[b*] | 20 | 0.01 | 0.33 | |

| | | | | |
|---|---|---|---|---|
| 6 | MeO-TPD@F4TCNQ[c] | 2 | 0.74 | 0.67 | |
| 7 | m-MTDATA@F4TCNQ[c] | 2 | 1.0 | 0.74 | |
| 8 | TPD@F4TCNQ[c] | 2 | 0.64 | 0.12 | |
| 9 | TCTA@F6TCNNQ[d] | 11 | 0.00 | 0.02 | |

Sources of experimental data: a [20] , b [21] ,c [22], d [23]

* This value was estimated from the ratio of the conductivities CBP and TCTA of Figure S2 of the Supporting information of the reference [b].

Comment: results for the materials from the reference [a] are relative values, which assumes that the ionization fraction of the BFDPB@CN6-CP is 1.0. Therefore we add the last column to the table representing the same normalization of the simulations of materials from [a] that is assuming that the ionization fraction is 100% for BFDPB@CN6-CP.

**Figure 2a** shows the correlation of the experimentally measured and simulated dopant ionization fractions, $\eta_{exp}$ and $\eta_{sim}$. **Figure 2b** shows the dopant ionization fraction versus the difference of the mean IP of the host (denoted as IP) and the mean of the EA of the dopant (denoted as EA), computed as described above.

Note that in the experimental work,[20] doping ionization ratio for BFDPB@CN6-CP was assumed to be 1.0, and three other ionization fractions from this study are put in relation to this value. However, it is unclear whether the ionization ratio is really 100% in this system (in experiment), which renders comparison between experiment and simulation difficult. For better comparison, we therefore applied the same scaling to the simulated result for materials *1-4*, depicted in the last column of **Table 1**. This cannot be applied to materials *6-8*, unfortunately, as no details on a reference material for 100% ionization ratio is given in the respective publications.[22] We find that the dopant ionization fraction decreases monotonously with increasing the offset energy $E_{off}$ = IP - EA in both experiment and simulation within each dataset, i.e. separately for material sets *1-4* and *6-8*. Further, applying the experimental scaling for simulations of materials *1-4*, we find a monotonous decrease with $E_{off}$ over the full dataset of all materials in the simulation results.

We stress that thorough analysis and the subsequent interpretation of data is rendered difficult by compiling experimental data from various sources. Future studies based on consistent experimental data may help to elucidate observed inconsistencies.

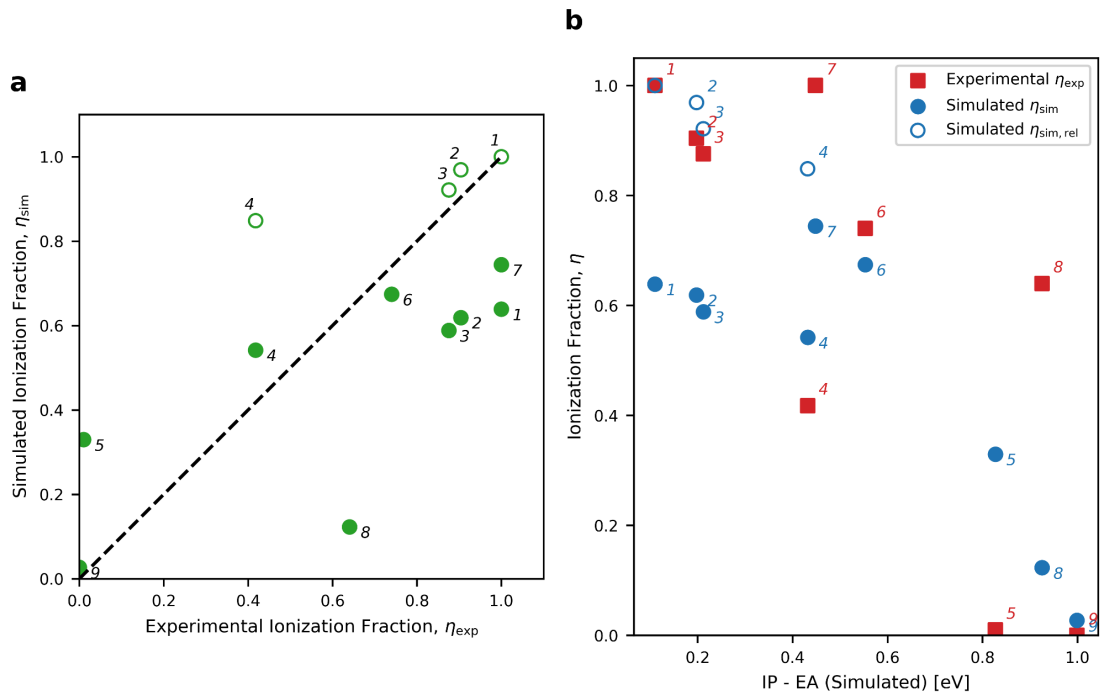

**Figure 2**. Experimental and simulated dopant ionization fraction: **a**. visualized as the cross-correlation plot. **b**. plotted versus simulated energy offset IP-EA. Materials are numbered according to **Table 1**. Hollow circles on both plots notify using relative rather than absolute value of ionization fraction, η$_{sim, rel}$, for materials *1-4*. See also the caption of **Table 1**.

### 2.1.2. Sensitivity of the dopant ionization fraction to the offset of the host IP and dopant EA

There is an apparent inconsistency between points *1–4* and points *6–8* (simulations) in our data, as the dopant ionization fraction exhibits a kink at a specific IP-EA difference, around $E_{off}$ = 0.4 eV, see **Figure 2b**. At this point, materials *4* and *7* have almost indistinguishable $E_{off}$, yet the difference in their ionized dopant fractions exceeds 20%. Since the sizes of the hosts and dopants are comparable, a steadily increasing η($E_{off}$) is anticipated as the $E_{off}$ decreases.

To investigate the deviation in the ionized dopant fraction between these materials, notably from different references [20] and [22], we systematically analyzed the impact of dopant EA on η using simulations. Specifically, for materials *4* and *7* we deliberately varied the electron affinity (EA) of the dopant molecule while keeping all other material properties exactly as computed with the multiscale workflow. We call these materials "fictitious" from now on.

**Figure 3a** shows the dependence of dopant ionization fraction of these fictitious materials (fictitious *4*, fictitious *7, solid lines*) along with actual materials, *4* and *7* (*solid circles*). The color maps the source where materials data are reported ([a] and [b] on the **Figure 3** correspond to the references [20] and [22], respectively). The results are intriguing: when the remaining data points from reference[20] (materials *1-3*) and reference[22] (materials *6,8*) are added to the plot (hollow circles), they aligned precisely with the line corresponding to the hypothetical variations of the $E_{off}$ in materials *4*[20] and *7*[22], respectively. This suggests that, to a large extent, the *exact molecular structures of the host and dopant do not significantly influence the fraction of ionized dopants*. In fact, the primary factor influencing η - along with the $E_{off}$ - is the dopant molar fraction: all materials from reference[20] are doped at 26 mol%, whereas those from reference[22] are doped at only 2%.

Previous simulations of model cubic morphologies have shown that in heavily doped materials, the ionization of weak dopants is enhanced while the ionization of the strong dopants is hindered by increasing doping concentration.[24] This same effect is observed here in the context of realistic, disordered morphologies. These findings explain the observed discrepancy between groups *1–4* and *7–8*. It also highlights the caution that has to be taken while comparing doping efficiency at different dopant concentrations. Trying to combine information from experiments and simulations, one can conclude that even in the worst-case scenario where the multi-scale model has an error of hundreds of meV, it is unlikely that dopants like BFDPB@CN6-CP from reference exhibit nearly 100% ionization.[20] Moreover, it is uncertain whether any of the materials measured in the experiments truly achieve full ionization. More systematic experimental works would be helpful to clarify this issue.

### 2.1.3. Host IP Estimation Using Ionization Fraction η

Given the observed correlation between ionization fraction η and IP-EA offset in simulations, measuring η can be utilized to estimate the host IP when the dopant EA levels are known. In this approach, dopants with known and varying strengths (i.e., different EA values) are employed, and η is measured with the goal of determining the host IP. The relationship $\eta_{sim}(E_{off})$, obtained from simulations, is nearly universal for a given dopant molar fraction (see **Supporting Figure 1**) and we can assume that $\eta_{sim}$= $\eta_{exp}$. Consequently, $E_{off}$ can be directly extracted from the systematic dependency of $\eta_{sim}(E_{off})$ at the respective concentration using the measured η for a given host, called η*. With information on dopant EA, this results directly in the host IP. In more details: from $\eta_{sim}(E_{off})$ we look up $E_{off}$ corresponding to the measured η*, which we call $E_{off}$*, and as we know the dopant EA, host IP = $E_{off}$* + EA.

To assess the feasibility of this approach, we evaluate its sensitivity, specifically identifying the required accuracy of η measurements to achieve chemical accuracy in determining the host IP. Chemical accuracy, defined as 1 kcal mol$^{-1}$ (0.043 eV), is commonly accepted as sufficient for practical applications. **Figure 3b** shows the derivative of the $E_{off}$ with respect to η. This derivative quantifies the sensitivity of $E_{off}$ to changes in η, serving as a measure of the method's robustness.

The results show that the sensitivity approaches 0.1 kcal/mol per percentage point, indicating that an experimental error of up to 10% in η measurement still yields an IP determination within chemical accuracy.

We note, nonetheless, that before this method can be reliably employed, it is necessary to resolve the ambiguity arising from measuring relative rather than absolute experimental ionization fractions.

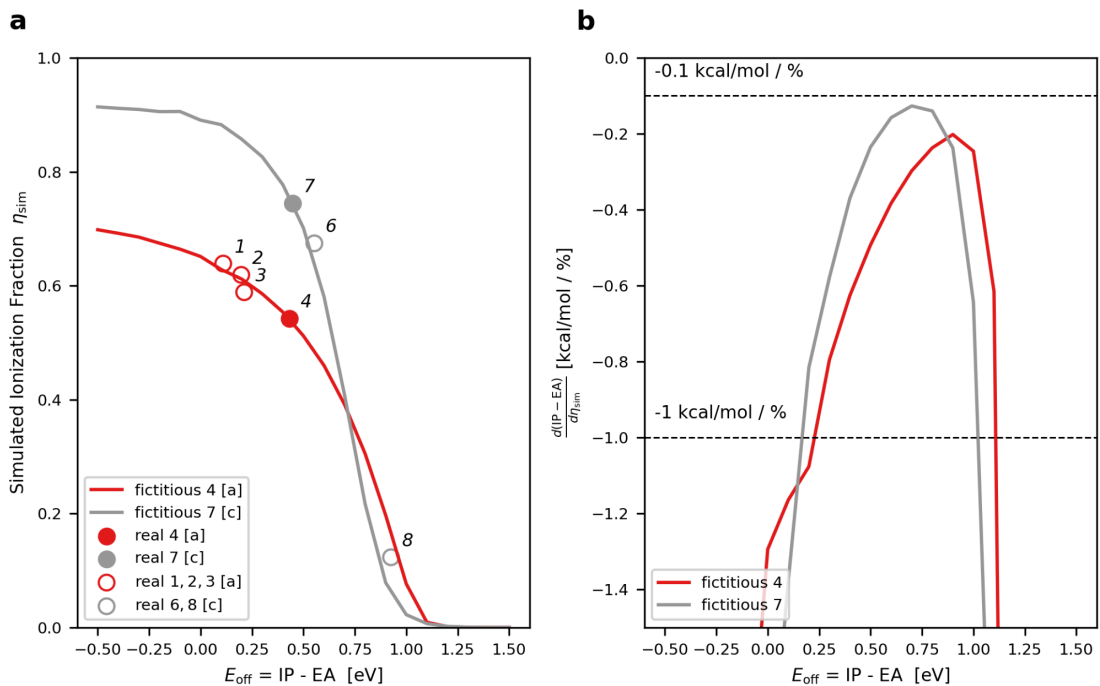

**Figure 3.** Influence of dopant molar fraction on the dopant ionization fraction and its sensitivity to the $E_{off}$ in fictitious TCTA@CN6-CP_0.26 (*4*) and m-MTDATA@F4TCNQ_0.02 (*7*) systems. **a**. Simulated ionization fraction $η_{sim}$ in real (circles) and fictitious (lines) materials as a function of the $E_{off}$. Fictitious materials are derived from the real materials *4* and *7* (filled circles) by deliberately changing EA of the dopant. Hollow circles represent the rest of materials from either ref. [a] or [c], i.e. dopant molar fractions are mapped by color. **b.** Sensitivity of the (IP - EA) offset to the ionization fraction $η_{sim}$ defined as the derivative of $E_{off}$ with respect to the $η_{sim}$, expressed in the chemical accuracy units (kcal mol$^{-1}$). In the best case, sensitivity nearly reaches 0.1 kcal mol$^{-1}$ per ionization percent, which implies that an experimental error of up to 10% in measuring of η can still result in an IP determination within chemical accuracy. References in the captions: a [20], c [22].

## 2.1.4. Role of Coulomb interactions

In the previous sections, we discussed the correlation of IP and EA and the dopant ionization fraction. However, to fully describe the charge transfer (CT) process, it is essential to include electrostatic interaction between the charged host and dopant molecules represented by their Coulomb interaction energy $V_C$.

Indeed, CT energy $E_{CT}$ represents the net energetic cost (or gain) associated with transferring an electron from a donor-like host (with ionization potential IP) to an acceptor-like dopant (with electron affinity EA) taking their Coulomb interaction $V_C$ into account:

$$E_{CT} = IP - EA + V_C,$$

However, $V_C$ remains difficult to measure directly, and its experimental determination was, to our knowledge, never reported for the doped organic materials. Consequently, most experimental studies report only IP and EA (onsets), and analyze ionization fraction dependence on the offset IP - EA, omitting $V_C$, which introduces uncertainty into our understanding of the dopant ionization process.

To overcome this uncertainty, we computed the $V_C$ as a function of the host-dopant center-of-mass (COM) distance, $d$ and the relative host-dopant orientation φ (φ is a general notion for multitude of parameters describing host-dopant mutual orientation introduced for convenience) by randomly selecting host-dopant pairs in the morphology in the relevant distances range. **Figure 4a** displays $V_C$ as a function of the inverse distance between host and dopant pairs, $d^{-1}$. For clarity, the direct distance is also indicated on the top axis. We find that for all materials the dependence of $V_C$ on $d^{-1}$ is comparable in terms of both *absolute values* and *trend*. At large distances (small $d^{-1}$), the dependence $V_C(d^{-1})$ is linear because the interaction between two distant and well-separated charge distributions is dominated by monopole-monopole interaction. At smaller distances, $V_C(d^{-1})$ reveals significant deviations from this classical monopole-monopole trend, indicating a phenomenon known as short-range overscreening.[10]

To allow for a systematic comparison across various doped materials, we introduce a characteristic $V_C$ called $V_{eff}$ at a distance that corresponds to the first peak of the host-dopant partial radial distribution function (RDF). This peak corresponds to the most probable distance from the dopant to the *nearest* host molecule, and - along with $V_{eff}$ - is shown at **Figure 4b**. For individual RDFs of all doped materials, see **Supporting Figure 2**. Interestingly, $V_{eff}$ varies across doped materials by not more than 175 meV. Using $V_{eff}$, the mean (rather than the individual molecule-pair specific) charge transfer energy (for the host-dopant distance equal to the first RDF peak) in the doped material can now be expressed as:

$$E_{CT} \text{(mean)} = IP\text{(mean)} - EA\text{(mean)} + V_{eff}.$$

**Figure 4c** shows the simulated ionization fraction $\eta_{sim}$ as a function of $E_{CT}$ for the fictitious material *based on* m-MTDATA@F4TCNQ (material *7*). Along with $\eta_{sim}$, we plot $f(\eta) = dE_{CT}/d\eta$, which is similar to the derivative plotted at **Figure 3b** but includes the important contribution $V_C$. The derivative $f(\eta)$ reaches its maximum at $E_{CT} \cong 0$. Comparing this result with **Figure 3b**, we see that the energy corresponding to the peak in $d(IP-EA)/d\eta$ aligns exactly with the characteristic Coulomb interaction energy introduced here, $-V_{eff}$.

These results on one hand justify that $V_{eff}$ is a good choice for the characteristic $V_C$. On the other hand, they also hint at another method for determining the IP of the host molecule. In this case, one would need a set of dopant molecules with known EA. To determine the ionization potential (IP) given a known Coulomb interaction energy $V_C$ and a variable electron affinity (EA), one can proceed as follows. Given the condition under which the charge transfer energy $E_{CT}$ vanishes, one can determine the IP of the given host as follows:

$$IP = EA^* - V_{eff}.$$

where the EA* is the EA of the dopant that satisfies the condition ($E_{CT} = 0$). $V_{eff}$ can be set to its typical value for instance 0.8 eV at least for materials similar to those used in this work, because its magnitude is very similar across doped materials. Interestingly, previous studies confirm weak dependence of $V_C$ on the host molecule;[25] even functionalizations that are heavily changing the microelectrostatic properties of a dopant molecule cannot change $V_C$ considerably.[10] Therefore, it must be possible to find a set of molecules with virtually indistinguishable $V_C$, but varying EA. Alternatively, one can compute the $V_C$ at the first RDF peak using the same method as used in this study, if higher accuracy is needed. Notably, since $V_C$ is a classical quantity - arising from the Coulomb interactions between the host and dopant charge densities - it is easier to compute accurately compared to IP or EA, which require higher-level quantum chemistry methods.

### 2.1.5. Discussion

The case studies show that the η relatively weakly depends on the morphology and dopant molecule structure (Figure 3a, Section 2.1.3). This means, the simulation results tolerate to an extent errors in morphology; and $V_C$ range is also relatively small across all systems (again, this means that the proposed methods tolerates errors in the $V_C$).

For the strong dopants, anions may be twice ionized leading to dianions.[26] This may lead to experimental \eta (which is dopant anion fraction) decreasing for very strong dopants (perhaps such dopants that IP - EA < -0.5 eV) because dianions rather than anions start prevailing. This effect generally has to be taken into account, but it will unlikely affect the position of the peak of the derivative $dE_{eff}/d\eta$, because the peak is characterized by dopants too weak to even ionize 100% dopants once. Therefore, possible dianion formation does not affect the accuracy of the proposed method.

We note that, by computing coulomb-interaction between all charges explicitly, the kMC model includes the effect of the so-called "dielectric catastrophe" reported recently.[8] Extraction of the dielectric permittivity that takes the increase of the polarization due to mobile ionized dopants into account is not discussed in this work. When referring to the relative dielectric permittivity in this study, we consider it as the permittivity of the material in the absence of mobile carriers generated as a result of doping.

It is known that the dopant-host mutual orientation leads to the variation of $V_C$ at a specific host-dopant distance ($V_C$ disorder) which may differ for different hosts.[10] However, the proposed method is mainly sensitive to the mean values of IP, EA and $V_C$, so that the magnitude of the $V_C$ disorder is not expected to significantly decrease the method accuracy.

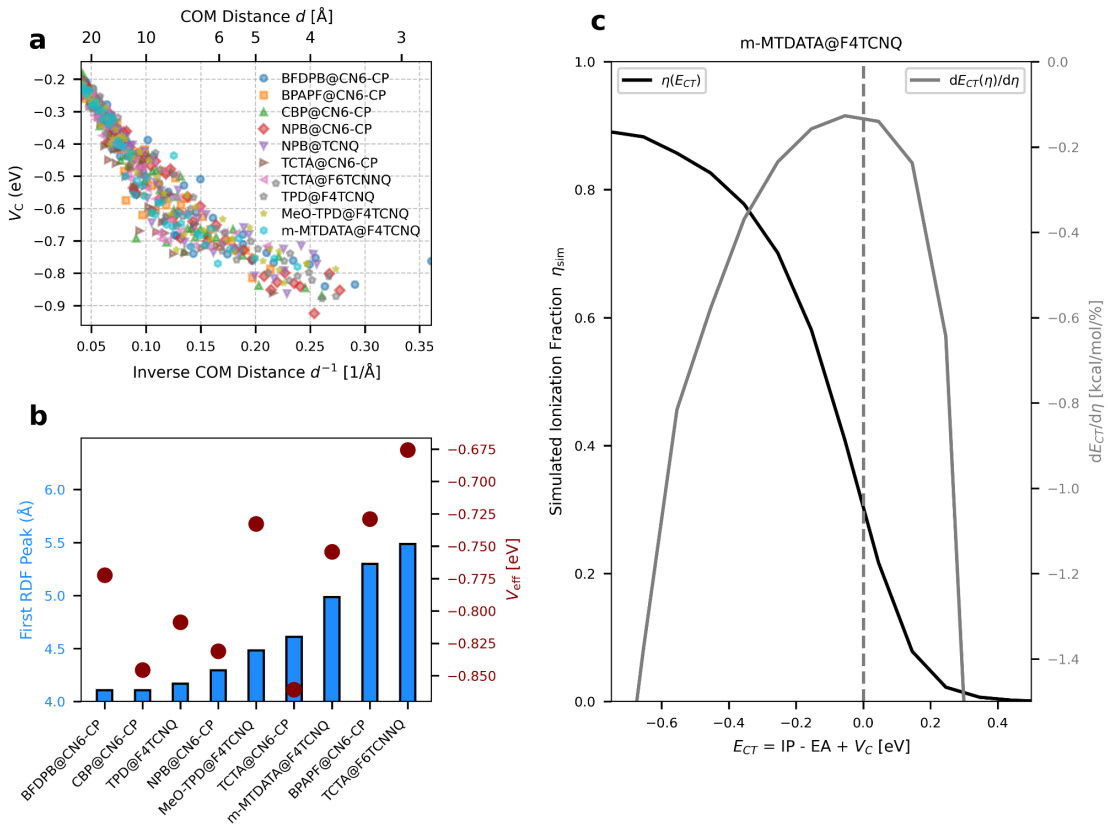

**Figure 4.** Effect of the host-dopant Coulomb interaction energy $V_C$ on the dopant ionization fraction. **a.** Host-dopant $V_C$ plotted vs. inverse/direct distance for representative host-dopant pairs showing similarity across all materials, including deviation from classical monopole-monopole interaction (short-range overscreening). **b.**

The first peak of the partial radial distribution function (RDF) and $V_C$ evaluated at this peak ($V_{eff}$). **c.** Simulated ionization fraction of the fictitious materials derived from m-MTDATA@F4TCNQ vs. the CT state energy $E_{CT}$, as well as the derivative of this dependence. At $E_{CT} = 0$ the derivative has a peak related to the resonance where IP - EA = -$V_C$, which can be used to determine IP of the host molecule if EA and $V_C$ are known or can be reliably estimated.

## 2.2. Doping-induced conductivity

Next, we explore insights to be gained from another well-established experimental technique: measuring the electrical conductivity of doped organic semiconductors (OSCs), which are commonly used to verify whether doping has been successful.

Conductivity in OSCs is proportional to the number of free carriers times the charge carrier mobility. As the number of free carriers in undoped small molecule organic thin films is negligible ($10^{12}$ - $10^{16}$ cm$^{-3}$) in comparison to doped OSCs, the conductivity of doped OSCs is expected to increase almost linearly with the number of free carriers created by doping. Although both superlinear and sublinear increases have been reported, [2,9,22] this detail is not critical to our subsequent discussion, which will focus on the much stronger exponential dependence of conductivity on the IP - EA offset, $E_{off}$.

In case of an activated doping fraction of a single digit percentage, the number of free charges would be increased to the order of $10^{19}$ cm$^{-3}$ for typical molecular volumes of ~ 1 nm$^3$ and hence increase conductivity by many orders of magnitude. Consequently one would roughly expect that the dependence of the conductivity σ on the charge transfer energy $E_{CT}$ can be described by a Fermi function:

$$\sigma = C\, E_F(E_{CT}), \quad (1)$$

where $E_F = 1/(1 + \exp(E_{CT}/k_B T))$ with $k_B$ and $T$ being the Boltzmann constant and the temperature, respectively; $C$ is the normalization constant.

Therefore, measuring conductivity of host materials doped with different (in terms of EA) dopants provides a robust check of the quality of measured or calculated charge transfer energies $E_{CT}$ and hence host IP and dopant EA.

**Figure 5** shows experimentally *measured* conductivity $\sigma_{exp}$ from various literature sources (as summarized in **Table 2)** over the *simulated* EA-IP difference. IP-EA offsets were either derived from experimental onset values (IP$_{onset}$ and EA$_{onset}$) or from computed mean values (simply IP and EA as before), depicted as blue crosses and red dots, respectively. Note that conductivity was reported for different concentrations for each material combination, ranging from 1% to 10%. As discussed above, dopant concentration has a significant impact on dopant ionization ratio and therefore conductivity, as visible in the noticeable spread along the vertical axis in **Figure 5**.

To understand the apparent trend of conductivity declining beyond a certain point, we next employ two methods to estimate the conductivity. The dashed black line in **Figure 5** depicts a function described by equation (1) with EA - IP as an argument and $V_C$ = 0.7 as fixed parameter used in $E_{CT}$, at a temperature of 300 K. Given the fact that for most materials relevant to this study, effective $V_C$ (called $V_{eff}$ above) is in the range of 0.7-0.85 eV, we assume this to be a robust and justified simplification of the model. A more systematic study where we used kinetic Monte-Carlo (kMC) simulations on the cubic lattice for a scan of EA - IP differences, with nearest-neighbours $V_C$ set to a fixed value of 0.7 eV (red solid line) aligns well with the estimation based on equation (1). For the kMC simulations we used a previously reported protocol (reference [9]) with improved treatment of host-dopant Coulomb interaction by setting $V_C$ to 0.7 eV for the nearest neighbors. Doping level in all model systems was set to 5 mol%. Other kMC simulation parameters are provided in **Methods** Section. We note that the prefactor of the function (1) is chosen so that the conductivity roughly matches the experimental one in case of 100 % dopant activation; in the kMC method, simulated conductivity was also normalized to match the experimental one in case of 100 % dopant activation.

We can see that the experimental data plotted versus computed IP-EA values (red dots) scatter around the simulated curves (lines), whereas measured values (blue crosses), i.e. $IP_{onset}$ - $EA_{onset}$ are off by up to 1 eV.

Possible reasons contributing to this discrepancy are:[16] 1) broadening of the density of HOMO/LUMO states due to the broadening of the density of states (energetic disorder). 2) relevant for charge transfer are adiabatic EA and IP values, whereas IP and EA derived via spectroscopical means are vertical IP and EA. 3) The onset states which are measured in UPS /IPS are rare. Since we have strongly localized electron-hole pairs, it is very rare to find a donor-acceptor pair in which both molecules are in the respective tail of the distribution, hence EAs and IPs closer to the mean contribute much more to the average activation probability. 4) The vibronic broadening which decreases IP onsets and increases EA onsets does not shift mean values. The net effect of both energetic disorder and relaxation amounts to 400 meV = 2x100 + 2x100 meV assuming 100 meV is a typical relaxation energy for IP/EA and the typical energetic disorder (see **Supporting Table 1** and **2**). The rest 600 meV is presumably due to broadening of the response functions in UPS/IPES due to the fact that both apparent UPS/IPES spectra are the convolution of the actual density of states with escape/dissipation functions. Although this additional 600 meV broadening seems very large, it aligns with the measured excessive disorder introduced by UPS/IPES.[27] Indeed, for P3HT the measured disorder of HOMO/LUMO from UPS/IPES spectrum is 255 meV/355 meV, while the apparent disorder of HOMO/LUMO from the Energy-Resolved Electrochemical Impedance Spectroscopy (ER-EIS) method - which, according to reference,[27] is claimed to map the intrinsic DOS without additional broadening - is 63/168 meV. The net excessive disorder of HOMO/LUMO as measured by UPS/IPES is thus 379 meV, which - being added to the mentioned 400 meV - is close to the typical difference between simulated (IP-EA) and experimental ($IP_{onset}$-$EA_{onset}$) observed in **Figure 5**.

We note that the excessive broadening of the peaks of the UPS as reported in reference[27] is much larger than the instrumental resolution of the state-of-the art UPS systems. The latter has been reported to be below 5 meV. This suggests that factors beyond instrumental resolution are responsible for the observed broadening in organic semiconductors

**Table 2**. Dopant molar fraction (from 1 mol% to 10 mol%) vs conductivity extracted from the literature.

| Material | Dopant Molar Fraction-Conductivity (mol%: S cm$^{-1}$) | Reference |
| --- | --- | --- |
| TCTA@F6TCNNQ | 10.00: 2.05x10$^{-8}$ | [a] |
| TCTA@CN6-CP | 8.76: 1.11x10$^{-5}$ | [b] |
| CBP@CN6-CP | 5.22: 2.68x10$^{-8}$ | [b] |
| MeO-TPD@F4TCNQ | 2.00: 1.00x10$^{-5}$ | [c] |
| MeO-TPD@F4TCNQ | 1.82: 1.36x10$^{-5}$<br>2.73: 2.54x10$^{-5}$<br>6.27: 1.39x10$^{-4}$ | [d] |
| MeO-TPD@F6TCNQ | 1.23: 1.50×10$^{-5}$<br>2.31: 2.86×10$^{-5}$<br>3.57: 9.92×10$^{-5}$<br>4.01: 7.74×10$^{-5}$<br>4.65: 1.17×10$^{-4}$<br>4.69: 1.32×10$^{-4}$<br>5.54: 1.24×10$^{-4}$<br>7.21: 1.96×10$^{-4}$ | [e] |
| BFDPB@F6TCNNQ | 1.97: 3.22×10$^{-6}$<br>3.85: 4.39×10$^{-6}$<br>5.87: 5.84×10$^{-6}$ | [e] |
| m-MTDATA@F4TCNQ | 2.00: 3.00x10$^{-7}$ | [c] |
| TPD@F4TCNQ | 2.00: 1.00x10$^{-7}$ | [c] |

Sources of experimental data: [a]: [23], [b]: [21], [c]: [22], [d]: [28], [e]: [29].

The main insight that we see at **Figure 5** is not only that the trend of the simulated conductivity matches well the experimental data (red line follows red points). The declination point of this dependence is around 0.7 eV, which is nothing more than the $V_C$ we set in the simulation for the nearest-neighbor $V_C$. We can thus use the simulated conductivity vs the IP-EA offsets for various EA as a method to evaluate the magnitude of host-dopant Coulomb interactions. The measured conductivity versus measured $IP_{onset}$ - $EA_{onset}$ data (marked by blue crosses), on the other hand, does not immediately allow for the extraction of $V_C$. Instead, the synergy of the experimental data and simulations is essential to evaluate $V_{eff}$.

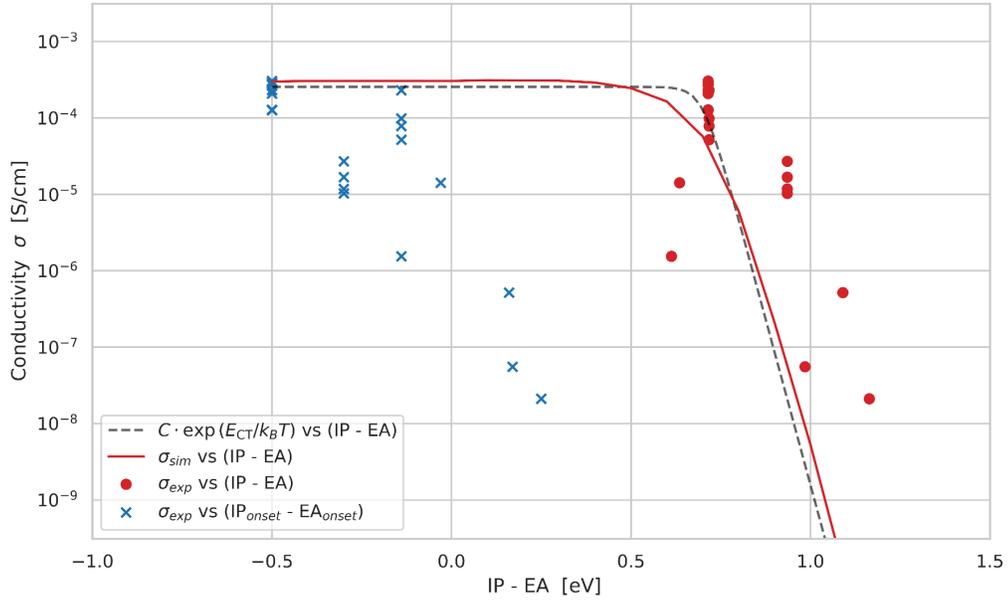

**Figure 5**: Conductivity in dependence of IP-EA difference. Red line: Simulated conductivity (kMC) for a cubic lattice system with 5% dopant concentration and $V_C$ = 0.7 eV for different IP-EA offsets. Dashed line: Fermi distribution of dopant activation according to formula (1) assuming $V_C$ = 0.7 eV. Blue crosses: measured conductivity and IP-EA offsets (from onsets) for various materials. Red dots: computed IP-EA offsets (from means) and *measured* conductivity for the same materials.

## 2. Conclusion

In conclusion, we have investigated how conductivity doping, traditionally employed to optimize charge transport and injection in organic electronic devices, can be repurposed as a tool to probe fundamental properties of organic materials, which are otherwise hard to access experimentally. Specifically, using the synergy of the experimental data and multiscale simulations, we have suggested a method to extract ionization potential and Coulomb interaction energy from the measured dopant ionization fraction and conductivity using the dopant molecules as agents. Practicality of the method to extract IP is evaluated: the chemical accuracy of 1 kcal mol$^{-1}$ of indirect measurement of the IP may require reasonable accuracy of the dopant ionization fraction measurement of 10%. Unlike spectroscopic methods like UPS and IPES, where onset IP /EA values are measured, such measurements on doped materials allows to determine the mean value of IP and the value of the coulomb interaction between host and dopant molecules. We note that this method is applicable for doped systems exhibiting integer charge transfer (ICT), which can be assumed for the systems at hand based on UV-vis-IR-spectra. For doped systems with orbital hybridization, adaptations would be required.[15]

## 3. Methods

### Morphology Generation

Single molecule conformations of all host and dopant materials were optimized using def2-TZVP@B3LYP level using the Parametrizer module.[19] Subsequently, the DihedralParametrizer module was used to compute molecule-specific forcefields for intramolecular interaction,[19] i.e. energy profiles for dihedral rotations and parameters for intramolecular Coulomb- and Lennard-Jones interaction. Resulting host and dopant molecular structures and force-fields were used to generate 3D morphologies of thin films with atomistic resolution for every host-dopant combination, with dopant molar fraction as specified in **Table 1,** by mimicking physical vapor deposition, as implemented in Deposit.[17] For each system, 3500 molecules were deposited on the rectangular box with 120x120 Angstroms base. Other simulation parameters of the Deposit method are identical to those specified in reference.[10]

### Electronic Properties (QuantumPatch)

Energetic disorder, IP and EA of the host and dopant molecules embedded into the corresponding doped material, and the host-dopant Coulomb interaction energies were computed using different modes of the quantum embedding method QuantumPatch as described below.[10,18]

### Energetic Disorder

The energetic disorder for host IP and dopant EA distributions materials was computed on the deposited morphologies using the QunatumPatch protocol with settings described in reference.[30] The results are summarized in **Supporting Table 1**.

### Ionization Potential (IP) and Electron Affinity (EA)

Mean IP of the host and mean EA of the dopant in the corresponding doped materials were computed using the QuantumPatch method with settings described in reference [18] for eight hosts and eight dopants per material for the dopant molar fractions as given in **Table 1**, if the dopant molar fraction is lower than 7 mol%, otherwise for the doped material with 7 mol%. For systems from **Table 2**, which are not in **Table 1**, and for which conductivity was reported at more than one dopant molar fraction between 1 and 10 mol%, IP and EA were computed using 7 mol% doping. In the dopant molar range from 1 to 10%, dependence of the IP of the host and EA of the dopant on the dopant molar fraction is negligible.

### Coulomb Interaction Energy $V_C$

For every system from **Table 1**, $V_C$ was computed for 50 host-dopant pairs embedded into the respecting doped material. The explicit polarization shell radius was set to 40 Angstroms, the implicit shell relative dielectric permittivity is set to 3.0. The rest of the settings are identical to those used in reference [10]. Same as for IP and EA simulations, dopant molar fraction was set to 7 mol%, if in reality it was lower than 7%.

## kMC Simulations (Lightforge)

### Dopant Ionization Fraction

Dopant ionization fraction is computed by simulating dopant ionization/deionization and charge carrier dynamics with kinetic Monte-Carlo method as implemented in Lightforge.[10] Molecules are represented as sites, with their positions defined by the centers of mass in the generated morphology. The system used in Lightforge is deduced from the generated morphology as follows. The original morphology as deposited by the Deposit protocol is first appended by its copies in the directions normal to the growth direction. The subset of the sites is taken, which are inside the cube in the center of this expanded geometry and the restriction that there are precisely 500 dopants inside this cube. Sources of the host IP and dopant EA: respecting quantum embedding simulations with QuantumPatch, yielding mean of the vertical IP and vertical EA distribution in the morphology. To obtain mean adiabatic IP and EA, relaxation energies computed in vacuum at B3LYP/def2-TZVP level (see **Supporting Table 2**), were added to the respecting bulk IP/EA values. The standard deviations of the IP/EA distribution are equal to hole/electron energetic disorder computed with QuantumPatch method as described above. Coulomb interaction energy

$V_C$ between host and dopant molecules, which are closer to one another than 25 Angstroms, are drawn from the explicitly computed distribution for each system as described above. Hole hops are allowed from a given site to 26 nearest sites; to accelerate achieving the equilibrium, all transfer integrals were set equal as described in reference [9].

## Conductivity

kMC simulations to compute the conductivity in the doped organic material were done on the simple *cubic* lattice of system volume 20x20x20 nm$^3$ with grid spacing of 1nm where 5 % of cites are assumed dopants, similarly as described in reference [9]. In contrast to the computation of dopant ionization ratio above, $V_C$ set to a fixed value of 0.7 eV for the next nearest neighbours. The simulations were done for a range of host IP / dopant EA differences. Dielectric permittivity was set to 3.0, Field strength was set to 50 mV/nm. Number of neighbors where the hop is allowed to is set to 26. Charge transfer rates are computed via Miller-Abrahams equation, where the prefactor is irrelevant as the conductivity was scaled to fit experimental data as described in Section 2.2. The temperature is 300 K. The simulation protocol is implemented in Lightforge.[9]

# Supporting Information

Supporting Information is available from the Wiley Online Library or from the author.

# Acknowledgments

This work was performed on the HoreKa supercomputer funded by the Ministry of Science, Research and the Arts Baden-Württemberg and by the Federal Ministry of Education and Research. The authors acknowledge support by the European Union via the HORIZON-EIC-2021-Transitionopen project DiaDEM (Grant Agreement No 101057564).

# Conflict of Interest Statement

W.W. holds shares of a KIT spinoff, Nanomatch GmbH, which markets software including methods developed by KIT. The remaining authors declare no competing interests.

# Data Availability Statement

The raw experimental data collected and analyzed during the current study are available in the GitHub repository at [https://github.com/NanomatchGmbH/doping_lib](https://github.com/NanomatchGmbH/doping_lib), specifically in the folder "experimental_data". The folder "simulations" contains the settings of the software modules used.

# References

[1]  H. Méndez, G. Heimel, A. Opitz, K. Sauer, P. Barkowski, M. Oehzelt, J. Soeda, T. Okamoto, J. Takeya, J.-B. Arlin, J.-Y. Balandier, Y. Geerts, N. Koch, I. Salzmann, *Angew. Chem. Int. Ed.* **2013**, *52*, 7751.
[2]  B. Lüssem, C.-M. Keum, D. Kasemann, B. Naab, Z. Bao, K. Leo, *Chem. Rev.* **2016**, *116*, 13714.
[3]  K. Walzer, B. Maennig, M. Pfeiffer, K. Leo, *Chem. Rev.* **2007**, *107*, 1233.
[4]  I. E. Jacobs, A. J. Moulé, *Adv. Mater.* **2017**, 1703063.
[5]  M. L. Tietze, J. Benduhn, P. Pahner, B. Nell, M. Schwarze, H. Kleemann, M. Krammer, K. Zojer, K. Vandewal, K. Leo, *Nat. Commun.* **2018**, *9*.
[6]  M. Schwarze, C. Gaul, R. Scholz, F. Bussolotti, A. Hofacker, K. S. Schellhammer, B. Nell, B. D. Naab, Z. Bao, D. Spoltore, K. Vandewal, J. Widmer, S. Kera, N. Ueno, F. Ortmann, K. Leo, *Nat. Mater.* **2019**, *18*, 242.
[7]  C. Gaul, S. Hutsch, M. Schwarze, K. S. Schellhammer, F. Bussolotti, S. Kera, G. Cuniberti, K. Leo, F. Ortmann, *Nat. Mater.* **2018**, *17*, 439.
[8]  M. Comin, S. Fratini, X. Blase, G. D'Avino, *Adv. Mater.* **2022**, *34*, 2105376.
[9]  A. Fediai, F. Symalla, P. Friederich, W. Wenzel, *Nat. Commun.* **2019**, *10*.
[10] J. Armleder, T. Neumann, F. Symalla, T. Strunk, J. E. Olivares Peña, W. Wenzel, A. Fediai, *Nat. Commun.* **2023**, *14*, 1356.
[11] A. D. Özdemir, S. Kaiser, T. Neumann, F. Symalla, W. Wenzel, *Front. Chem.* **2022**, *9*.
[12] A. Fediai, T. Neumann, F. Symalla, T. Strunk, *SID Symp. Dig. Tech. Pap.* **2024**, *55*, 482.
[13] Y. Zeng, G. Han, Y. Yi, *Sci. China Chem.* **2024**, *67*, 3675.
[14] I. Salzmann, G. Heimel, S. Duhm, M. Oehzelt, P. Pingel, B. M. George, A. Schnegg, K. Lips, R.-P. Blum, A. Vollmer, N. Koch, *Phys. Rev. Lett.* **2012**, *108*.
[15] I. Salzmann, G. Heimel, M. Oehzelt, S. Winkler, N. Koch, *Acc. Chem. Res.* **2016**, *49*, 370.
[16] F. Symalla, A. Fediai, T. Neumann, T. Strunk, *SID Symp. Dig. Tech. Pap.* **2024**, *55*, 607.
[17] T. Neumann, D. Danilov, C. Lennartz, W. Wenzel, *J. Comput. Chem.* **2013**, *34*, 2716.
[18] J. Armleder, T. Strunk, F. Symalla, P. Friederich, J. Enrique Olivares Peña, T. Neumann, W. Wenzel, A. Fediai, *J. Chem. Theory Comput.* **2021**, *17*.
[19] T. Neumann, T. Strunk, A. Feidai, F. Symalla, *SID Symp. Dig. Tech. Pap.* **2023**, *54*, 822.
[20] B. Nell, K. Ortstein, O. V. Boltalina, K. Vandewal, *J. Phys. Chem. C* **2018**, *122*, 11730.
[21] Y. Liu, B. Nell, K. Ortstein, Z. Wu, Y. Karpov, T. Beryozkina, S. Lenk, A. Kiriy, K. Leo, S. Reineke, *ACS Appl. Mater. Interfaces* **2019**, *11*, 11660.
[22] M. Pfeiffer, K. Leo, X. Zhou, J. S. Huang, M. Hofmann, A. Werner, J. Blochwitz-Nimoth, *Org. Electron.* **2003**, *4*, 89.
[23] F. Zhang, A. Kahn, *Adv. Funct. Mater.* **2018**, *28*, 1703780.
[24] A. Fediai, A. Emering, F. Symalla, W. Wenzel, *Phys. Chem. Chem. Phys.* **2020**, *22*, 10256.
[25] Y. Zeng, G. Han, Y. Yi, *Sci. China Chem.* **2024**, *67*, 3675.
[26] D. Kiefer, R. Kroon, A. I. Hofmann, H. Sun, X. Liu, A. Giovannitti, D. Stegerer, A. Cano, J. Hynynen, L. Yu, Y. Zhang, D. Nai, T. F. Harrelson, M. Sommer, A. J. Moulé, M. Kemerink, S. R. Marder, I. McCulloch, M. Fahlman, S. Fabiano, C. Müller, *Nat. Mater.* **2019**, *18*, 149.
[27] H. Bässler, D. Kroh, F. Schauer, V. Nádaždy, A. Köhler, *Adv. Funct. Mater.* **2021**, *31*, 2007738.
[28] S. Olthof, W. Tress, R. Meerheim, B. Lüssem, K. Leo, *J. Appl. Phys.* **2009**, *106*, 103711.
[29] T. Menke, Molecular Doping of Organic Semiconductors: A Conductivity and Seebeck Study. Dissertation, TU Dresden, Dresden, **2013**.
[30] S. Kaiser, T. Neumann, F. Symalla, T. Schloder, A. Fediai, P. Friederich, W. Wenzel, *Front. Chem.* **2021**, *9*.

# ToC figure

A. Fediai*, F. Symalla, T. Neumann, W. Wenzel

**Using dopants as agents to probe key electronic properties of organic semiconductors**

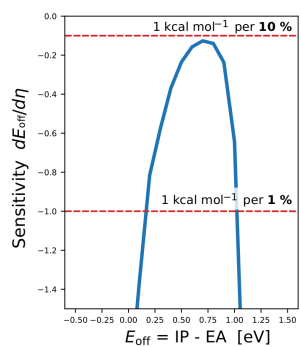

Dopants are typically used in organic electronics to enhance conductivity, but here we demonstrate their potential as probes for fundamental material properties. By integrating experimental data and multiscale simulations, we show how dopant ionization and conductivity measurements enable accurate extraction of ionization potential and Coulomb interaction energy, crucial for OLED and OPV design.